\documentclass[a4paper,UKenglish,cleveref, autoref, thm-restate]{lipics-v2021}

\usepackage{booktabs}
\usepackage{tcolorbox}
\tcbuselibrary{skins,breakable}
\usepackage{enumitem}
\usepackage{fontawesome5}
\usepackage{subcaption}

\newtcolorbox{rqbox}[2][]{
    enhanced,
    breakable,
    colback=#2!5!white,
    colframe=#2!60!black,
    colbacktitle=#2!60!black,
    coltitle=white,
    fonttitle=\bfseries\small,
    title={#1},
    boxrule=0.5pt,
    arc=2pt,
    left=4pt,
    right=4pt,
    top=2pt,
    bottom=2pt,
    titlerule=0pt,
    toptitle=2pt,
    bottomtitle=2pt,
}

\newenvironment{rqlist}{
    \begin{itemize}[
        leftmargin=1.4em,
        itemsep=2pt,
        topsep=2pt,
        label={\small\faHandPointRight}
    ]
}{
    \end{itemize}
}

\title{Predicting LLM Performance from Prompt Linguistic Features: An Empirical Study in Requirements Engineering} 

\titlerunning{Linguistic Properties of Prompts as Predictors of LLM Performance} 
\author{Quim Motger}{Universitat Polit\`ecnica de Catalunya, Barcelona, Spain}{joaquim.motger@upc.edu}{}{}
\author{Alessio Miaschi}{ItaliaNLP Lab, Istituto di Linguistica Computazionale ``A. Zampolli'' (CNR-ILC), Pisa, Italy}{}{}{}
\author{Xavier Franch}{Universitat Polit\`ecnica de Catalunya, Barcelona, Spain}{}{}{}
\author{Mohammad Amin Zadenoori}{University of Padua, Padua, Italy}{}{}{}
\author{Alessio Ferrari}{University College Dublin, Dublin, Ireland}{}{}{}
\authorrunning{Q. Motger, A. Miaschi, X. Franch, M.~A. Zadenoori, and A. Ferrari}

\Copyright{Quim Motger, Alessio Miaschi, Xavier Franch, Mohammad Amin Zadenoori, and Alessio Ferrari}
\hideLIPIcs

\nolinenumbers



\relatedversion{} 

\begin{document}

\maketitle

\begin{abstract}
\textbf{Background.} LLM outputs are highly sensitive to prompt formulation. Small changes in wording or phrasing can lead to substantial differences in output quality. 
This sensitivity is particularly important in software engineering, where prompts guide tasks such as requirements analysis, code generation, and artifact synthesis in general.  Ineffective prompt formulations can produce unreliable artefacts, with consequences for both task outcomes and the broader development process. 
However, practitioners lack principled ways to assess prompt formulation before inference (i.e., LLM execution), making prompt selection dependent on resource-intensive LLM calls and trial-and-error refinement. 
\textbf{Aims.} We investigate whether measurable linguistic properties of prompts can predict LLM performance before inference, thereby supporting low-cost prompt selection and refinement. We validate this hypothesis in the context of binary requirements classification by targeting performance metrics (F1, F2, precision, recall). 
\textbf{Method.} We generate 9,000 linguistically controlled prompt variants derived from 100 initial prompts by varying 30 linguistic metrics. Variants are evaluated using five open-source LLMs on 625 requirements annotated with ground-truth labels. Regression predictors are trained using stratified 10-fold cross-validation with permutation-based significance testing. Feature importance analysis identifies cross-LLM and model-specific linguistic predictors.
\textbf{Results.} Linguistic features significantly predict prompt performance across all targets ($R^2 \in [0.38, 0.42]$, $q < 0.05$). 
Syntactic and morphosyntactic features drive most of the predictive signal. 
Cross-LLM consistent predictors include compound dependency distribution, conjunction density, and raw-text indicators such as word and sentence length, reflecting sensitivity to domain-specific vocabulary and complex structures. 
\textbf{Conclusions.} 
These results draw practical observations for prompt engineering, including the overlap between linguistic patterns that reduce LLM performance and those that increase human comprehension difficulty, as well as the irrelevance of lexical variety as a prompt-quality dimension. 
More broadly, they show that linguistic profiling combined with standard regression models provides an effective, interpretable signal for prompt characterisation, applicable as a low-cost prior before engaging costly optimisation pipelines.
\end{abstract}

\section{Introduction}
\label{sec:introduction}

The increasing adoption of large language models (LLMs) in software engineering (SE) has demonstrated great potential to automate a wide range of tasks~\cite{Hou2024}, from code generation~\cite{Wang2025} to requirements analysis~\cite{Ronanki2024}. 
However, LLM outputs are highly sensitive to prompt formulation: small changes in wording, structure or phrasing can lead to substantial differences in output quality~\cite{errica-etal-2025-wrong}. 
This effect has been documented across task types and model families, from accuracy swings~\cite{sclar2024quantifyinglanguagemodelssensitivity} to inconsistent model rankings~\cite{mizrahi-etal-2024-state} under prompt formatting changes. 
The consequences are particularly acute in SE. While reproducibility, interpretability of prompt artefacts, and developer trust are central concerns, even binary judgements have been shown to flip under minor paraphrasing~\cite{Han2026}.
This sensitivity makes prompt design a critical --- yet poorly understood --- determinant of task performance, with no established principles to guide SE practitioners in formulating effective prompts~\cite{Ronanki2025}.

Prompt engineering has emerged as the community's response to this challenge, aiming to formulate prompts that elicit reliable, high-quality outputs from LLMs.
In SE and requirements engineering (RE), this work has so far been predominantly structural or template-based, with researchers proposing design frameworks~\cite{de2025framework}, role and context patterns~\cite{khojah2025impact}, and task-specific templates for code generation~\cite{shin2025prompt,bruni2025benchmarking}, test generation~\cite{ouedraogo2026prompt}, and requirements classification~\cite{binkhonain2025prompts}.
Overall, systematic reviews confirm the centrality of structural design choices to LLM performance in these settings~\cite{huang2025prompt}.
A complementary line of work applies automatic prompt engineering (APE), which frames prompt design as an optimisation problem and searches for high-performing prompts through iterative rewriting, meta-prompting, or reinforcement learning~\cite{zhou2023largelanguagemodelshumanlevel,kong-etal-2024-prewrite,ye-etal-2024-prompt,Liu2024,taherkhani2024automated,Zadenoori2025}.
Yet both directions have limitations: structural and template-based approaches typically prescribe a fixed design and do not account for how small linguistic variations within that design affect output, while APE requires repeated LLM calls to evaluate candidate prompts, making it computationally expensive and energy-intensive~\cite{Rubei2025}. What is missing is a lightweight way to estimate prompt effectiveness from linguistic properties of the prompt itself, before inference. Such a signal would complement both directions by constraining the search space, guiding prompt generation, and supporting the selection of promising prompts without running the LLM at every step.

We therefore hypothesise that \textbf{measurable linguistic properties of prompts are systematic predictors of LLM performance}. If such relationships can be modelled, prompt designers can assess and improve prompts analytically before engaging costly optimisation pipelines, reducing the number of LLM interactions required during prompt development.

To validate this hypothesis, a controlled empirical study is required in a setting where (i) prompt variability can be systematically manipulated,  (ii) ground-truth performance can be reliably measured, and (iii) a sufficient volume of prompt variants can be generated and evaluated. 
Requirements engineering (RE) offers a particularly suitable instantiation: RE tasks such as requirements classification are well-defined, supported by established benchmarks, and have been shown to benefit from LLM-based automation~\cite{zadenoori2025largelanguagemodelsllms}. 
Accordingly, the goal of this study is \textbf{to validate the above hypothesis in the context of binary requirements classification}, using a controlled experimental setting in which linguistic properties of prompts are systematically manipulated and their effect on performance --- focusing on functional correctness as defined by ISO/IEC~25010~\cite{iso25010} --- is measured and modelled.

Our study makes the following contributions:

\begin{itemize}
    \item \textbf{A prompt variability framework} that combines controlled paraphrasing --- guided by targeted manipulation of linguistic metrics --- with systematic regression-based analysis of the impact of linguistic properties on LLM performance.
    \item \textbf{A benchmark dataset} of 9,000 linguistically diverse prompt variants for binary requirements classification, evaluated against five open-source LLMs and annotated with performance metrics.
    \item \textbf{A ranked assessment of linguistic predictors} of prompt performance, distinguishing features that generalise across LLMs from those that are model-specific, providing actionable guidance for prompt engineers targeting heterogeneous model deployments.
\end{itemize}

The remainder of this paper is structured as follows. Section~\ref{sec:background} reviews relevant background on prompt engineering and linguistic profiling. Section~\ref{sec:method} describes the research 
method. Section~\ref{sec:results} presents the results. Section~\ref{sec:discussion} discusses findings and implications. Section~\ref{sec:ttv} addresses threats to validity. Section~\ref{sec:rel-work} surveys related work. Section~\ref{sec:conclusions} concludes the paper.

\section{Background}
\label{sec:background}

\subsection{Prompt Engineering}

Prompt engineering is the process of designing and refining prompts --- i.e., textual inputs or queries given to LLMs --- to maximise the performance of the model on specific tasks~\cite{marvin2023prompt,Zadenoori2025}. More formally, prompt engineering can be cast as an optimisation problem~\cite{ye-etal-2024-prompt,Zadenoori2025}, where the objective is to identify the optimal prompt $p^*$ that maximises a task-specific evaluation metric over a ground-truth dataset $D = \{(x_i, y_i)\}_{i=1}^{n}$, using an LLM $\mathcal{M}$:

\begin{equation}
    p^* = \arg\max_{p} \ f\left(\{\mathcal{M}(x_i, p)\}_{i=1}^{n}, \{y_i\}_{i=1}^{n}\right)
\end{equation}

\begin{itemize}
    \item $D = \{(x_i, y_i)\}_{i=1}^{n}$ is the ground-truth dataset, where each $x_i$ is an input instance and each $y_i$ is its corresponding ground-truth output;
    \item $\mathcal{M}$ is the LLM, which takes a prompt $p$ and an input $x_i$ to produce a predicted output $\hat{y}_i = \mathcal{M}(x_i, p)$. The prediction $\hat{y}_i$ is expected to match, or approximate, the corresponding ground-truth output $y_i$;
    \item $f$ is the task-specific evaluation function to be maximised. It compares the predictions induced by prompt $p$, namely $\{\hat{y}_i\}_{i=1}^{n}$, with the ground-truth outputs $\{y_i\}_{i=1}^{n}$, and returns a scalar score indicating the effectiveness of the prompt execution. 
\end{itemize}

In the context of requirements classification, each $x_i$ represents a textual requirement, and each $y_i$ denotes its manually annotated class. For example, a requirement may be classified as functional, as in ``\emph{When the user submits the form, the system shall validate the input}'', or as non-functional, when it concerns quality attributes such as security, performance, or others, as in ``\emph{The system shall respond within two seconds under normal load}''. In this setting, the evaluation function $f$ can be instantiated using a classification metric, such as accuracy or F1, to measure how well the predicted requirement classes match the manually annotated ones.

Prompt engineering can be performed either manually or automatically, through APE approaches~\cite{zhou2023largelanguagemodelshumanlevel}. In manual prompt engineering, human experts craft and refine prompts based on their domain knowledge and on iterative evaluation against a subset of the ground-truth data. In APE, instead, the search for effective prompts is partially or fully automated, with candidate prompts being generated and evaluated according to $f$.


\subsection{Linguistic Profiling}

Linguistic profiling is the methodology of characterising a text through a structured set of measurable linguistic features extracted from automatic annotation. Originally developed for stylometric and authorship attribution tasks~\cite{van2004linguistic}, the approach has since been extended to a broad range of NLP applications, including readability assessment~\cite{dell2011read}, native language identification~\cite{malmasi2018native}, and, more recently, the evaluation and interpretation of neural language models~\cite{miaschi-etal-2020-linguistic,miaschi-etal-2024-evaluating}.

The central idea is that texts can be represented as vectors of quantitative descriptors computed over multiple levels of linguistic annotation~---~typically \emph{raw text} (e.g., document length, average word and sentence length), \emph{lexical} (e.g., lexical density, type/token ratio), \emph{morphosyntactic} (e.g., distribution of parts of speech and inflectional categories), and \emph{syntactic} (e.g., dependency relation distributions, parse tree depth). These features capture stable, interpretable properties of linguistic structure rather than task-specific surface cues, making them well-suited to comparative analyses across texts, authors, genres, or models.

More formally, given a text $t$, linguistic profiling produces a feature vector
\begin{equation}
\phi(t) = \langle f_1(t), f_2(t), \ldots, f_n(t) \rangle \in \mathbb{R}^n
\end{equation}
where each $f_i$ is a function that maps the text~---~typically after automatic linguistic annotation (tokenisation, POS tagging, dependency parsing)~---~to a real-valued descriptor. The annotation step is commonly based on standardised formalisms such as Universal Dependencies~\cite{de2021universal}, which provide cross-linguistically consistent representations of morphosyntactic and syntactic structure. The resulting vector $\phi(t)$ can then be used as input to downstream analyses, including classification, regression, correlation studies, or distributional comparison between groups of texts.

Two properties make linguistic profiling well-suited to the present study. First, extracted features are \emph{interpretable}: each descriptor maps to a well-defined linguistic property, linking predictive analysis to specific textual characteristics. Second, features are \emph{computed deterministically} from raw text --- no learned parameters, no LLM inference --- making them inexpensive to compute and reproducible. 
Hence, we adopt linguistic profiling to characterise prompts as structured linguistic artefacts. By representing each prompt variant as a feature vector $\phi(p)$ over a comprehensive set of linguistic descriptors, we enable the systematic regression-based analysis of how linguistic structure relates to LLM performance. 

\section{Research Method}
\label{sec:method}

\subsection{Design}

Following the ABC framework for SE research~\cite{stol2018abc}, this work is classified as a \textbf{knowledge-seeking empirical study} combining two research strategies: a \textbf{sample study}, in which a large population of prompt variants is analysed to uncover statistical relationships between linguistic features and performance metrics, and a \textbf{laboratory 
experiment}, in which prompt variants are evaluated in a controlled 
environment using multiple LLMs.

Two research questions guide the study:

\begin{itemize}
    \item \textbf{RQ1}: To what extent can the performance of LLM-based requirements binary classification be predicted from linguistic features of prompts?
    \item \textbf{RQ2}: Which linguistic features show the highest predictive relevance in LLM-based requirements binary classification?
\end{itemize}

\textbf{RQ1} addresses \emph{predictive feasibility}: whether linguistic features of prompts contain sufficient information to predict performance outcomes. It is answered through stratified 10-fold cross-validation across four performance targets (Precision, Recall, F1, F2) using five regression models, producing predictive performance scores that indicate whether linguistic features are a viable predictor of prompt performance.

\textbf{RQ2} addresses \emph{linguistic predictor relevance}, and is answered through permutation-based feature importance analysis on the best-performing regression setup from RQ1, compared across LLMs and performance targets. This analysis identifies which linguistic properties contribute most strongly to prompt performance prediction, both at the level of individual features and feature groups. It also examines whether highly relevant features exhibit stable patterns across LLMs or whether their relevance varies depending on the target model.

Both RQs are framed as joint predictive analyses rather than per-feature statistical tests. Linguistic profiling produces a high-dimensional, correlated feature space in which classical feature-wise testing is hard to interpret and exposed to multiple-comparison issues. Regression with cross-validation and permutation-based significance testing instead evaluates whether the feature set carries reliable joint predictive signal, while subsequent feature-importance analysis recovers the interpretable patterns within that signal. The study therefore combines predictive utility with explanatory insight.

\begin{figure}[t]
    \centering
    \includegraphics[width=\columnwidth]{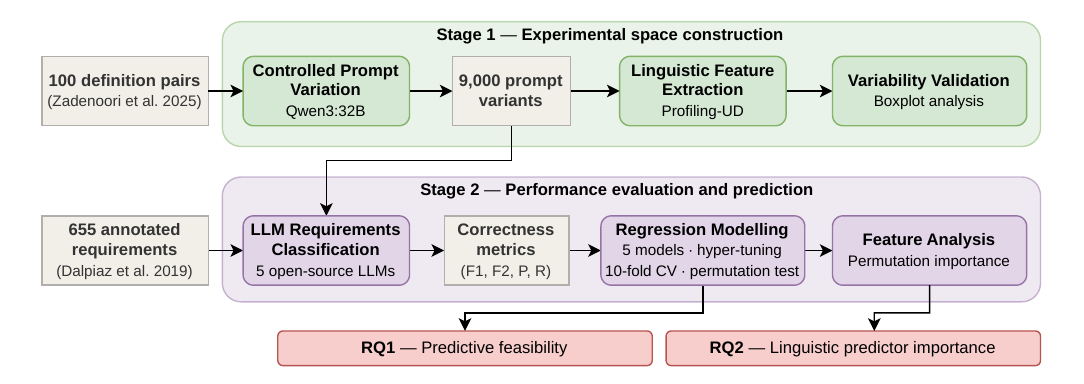}
    \caption{Overview of the research methodology.}
    \label{fig:overview}
\end{figure}

To support this analysis, the methodology is structured into two stages, illustrated in Figure~\ref{fig:overview}. 
\textbf{Stage~1} constructs and validates the experimental space: it generates linguistically controlled prompt variants through targeted paraphrasing guided by a set of linguistic metrics, and verifies that the manipulation effectively increases linguistic variability across the prompt population. 
\textbf{Stage~2} evaluates each prompt variant against multiple open-source LLMs on the binary requirements classification task, collecting performance metrics, and uses the resulting prompt--performance pairs to train regression models and analyse feature importance to answer the research questions.

\subsection{Stage 1: Experimental Space Construction}

\subsubsection{Task and Dataset}

The target task is \emph{binary requirements classification}: given a software requirement expressed in natural language, determine whether it is \emph{Functional} (F) or \emph{Non-functional} (NF). This is one of the most studied and well-understood tasks in the field of Natural Language Processing (NLP) for RE~\cite{Zhao2021}, supported by established benchmarks~\cite{Motger2026}, making it particularly suitable for controlled experimentation where precise performance measurement is required.

We evaluate the task on the PROMISE-NFR dataset~\cite{ClelandHuang2007}, the most popular dataset on requirements classification~\cite{Motger2026}. The dataset provides \emph{625 annotated software requirements} derived from the PROMISE repository~\cite{ClelandHuang2007}.
For the prompt definitions, we reuse the dataset introduced by Zadenoori et al.~\cite{Zadenoori2025}, which provides 100 few-shot prompt variants based on \emph{alternative natural language definition pairs} for the classes F and NF. These definitions vary in wording and phrasing while referring to the same underlying concepts, making them a natural starting population for linguistic manipulation. We focus on few-shot prompting as it has been shown to reduce LLMs' sensitivity to prompt formulation~\cite{zhuo-etal-2024-prosa}.


The prompt structure itself --- system context, examples, and classification instruction --- is kept constant across all variants (Listing~\ref{lst:prompt}). Linguistic manipulation is applied exclusively to the requirements class definitions. 


This design choice reflects a deliberate trade-off between \emph{breadth} and \emph{depth}. Rather than varying task, dataset, or prompting strategy (e.g., zero-shot, few-shot, Chain-of-Thought) simultaneously, we fix these dimensions and conduct an in-depth empirical analysis of linguistic variability within a single, well-controlled setting. This is consistent with established practice in empirical SE, where studies focusing on a single task and benchmark are a recognised and productive approach to validating novel hypotheses before broader generalisation is attempted~\cite{Felizardo2024,Cui2025,Salman2025,Rahman2025}. If measurable linguistic properties of prompts prove predictive under these controlled conditions --- across a large number of prompt variants, multiple linguistic metrics, and five LLMs --- the hypothesis gains the empirical support needed to motivate replication across other RE tasks, datasets, and prompt strategies in future work.

\begin{lstlisting}[
caption={Prompts used for requirements classification (adapted from Zadenoori et al. \cite{Zadenoori2025})},
label={lst:prompt},
breaklines=true,
basicstyle=\ttfamily\footnotesize
]
SYSTEM PROMPT
-------------
As an expert system for classifying software requirements, your job is to carefully review each requirement and place it into one of these two classes: Functional, Non-functional
USER PROMPT
-----------
Definitions:

Functional: "{functional_def}"
Non-functional: "{non_functional_def}"

Examples:

Functional:
- The system shall allow modification of the display.
- The system shall offer a display of all the Events in the exercise.

Non-functional:
- 90% of untrained realtors shall be able to install the product
  without instructions.
- The product is expected to run on Windows CE and Palm operating
  systems.

Requirement:
"{text}"

Using the Definitions above, classify the requirement and provide the
final label in the format:

Label: [Your Class Label Here]
\end{lstlisting}

\subsubsection{Linguistic Feature Extraction}
\label{sec:lin-feat-ex}

\begin{table}[t]
\centering
\caption{Linguistic feature groups extracted using the Profiling-UD 
tool~\cite{brunato2020profiling}.}
\label{tab:features}
\small
\begin{tabular}{p{2.5cm} p{5.3cm} c p{4cm}}
\toprule
\textbf{Group} & \textbf{Description} & \textbf{\#} & 
\textbf{Example feature} \\
\midrule
Raw Text  & 
Surface-level statistics on document, sentence, and word length. & 
4 & 
\texttt{tokens\_per\_sent}: avg.\ number of tokens per sentence \\
\addlinespace
Lexical Variety & 
Type/Token Ratio computed over standardised windows of 100 and 200 tokens, for both lemma and word form. & 
6 & 
\texttt{ttr\_lemma\_chunks\_100}: lexical variety over first 100 tokens \\
\addlinespace
Morphosyntactic Information & 
Percentage distribution of the 17 Universal POS categories and overall lexical density. & 
33 & 
\texttt{upos\_dist\_VERB}: proportion of main verbs in the text \\
\addlinespace
Inflectional Morphology & 
Distribution of verbal inflection features: tense, mood, form, gender, number, and person. & 
14 & 
\texttt{verbs\_tense\_dist\_Pres}: proportion of present-tense verbs \\
\addlinespace
Verbal Predicate Structure & 
Frequency and arity of verbal heads, including distribution of verbs by number of instantiated dependency links. & 
9 & 
\texttt{avg\_verb\_edges}: avg.\ number of dependents per verbal head \\
\addlinespace
Global \& Local Parse Tree Structures & 
Syntactic tree depth, clause length, dependency link length, and prepositional chain depth. & 
9 & 
\texttt{avg\_max\_depth}: avg.\ maximum depth of dependency trees \\
\addlinespace
Order of Elements & 
Distribution of pre- and post-verbal positions of subjects and objects. & 
4 & 
\texttt{subj\_pre}: proportion of subjects preceding the verb \\
\addlinespace
Syntactic Relations & 
Percentage distribution of the 37 Universal Dependencies relation types. & 
37 & 
\texttt{dep\_dist\_nsubj}: proportion of nominal subject dependencies \\
\addlinespace
Use of Subordination & 
Distribution, position, and embedding depth of subordinate clauses. & 
8 & 
\texttt{sub\_proposition\_dist}: proportion of subordinate clauses \\
\bottomrule
\end{tabular}
\end{table}

To characterise linguistic properties, we adopt the linguistic profiling framework proposed by Brunato et al.~\cite{brunato2020profiling}. In particular, we extract features using the \textbf{Profiling-UD} tool~\cite{brunato2020profiling}, a tool that allows the extraction of 124 properties representative of the linguistic structure underlying a sentence and derived from raw, morphosyntactic and syntactic levels of annotation based on the UD formalism \cite{de2021universal}. These features have been shown to play a highly predictive role when leveraged by traditional learning models on various classification problems and can also be effectively used to profile the knowledge encoded in the internal representations of Language Models and to evaluate their abilities in following specific linguistic constraints \cite{miaschi-etal-2020-linguistic,miaschi-etal-2024-evaluating}. Table~\ref{tab:features} summarises the nine feature groups extracted by Profiling-UD, together with the number 
of features per group and a representative example.



For controlled prompt generation, we select \textbf{30 linguistic metrics} spanning four groups: raw text, parse tree structure, verbal predicate structure, and subordination. 
These groups were chosen because they operate at distinct levels of linguistic description --- surface, syntactic, and clausal --- and have been shown to contribute independently and non-redundantly to text characterisation tasks such as readability assessment and genre classification~\cite{brunato2020profiling, miaschi-etal-2020-linguistic}. 
Manipulating metrics across these four dimensions therefore ensures that the generated prompt variants differ along genuinely distinct linguistic axes, rather than along correlated surface properties that would reduce the effective diversity of the experimental space. 
The complete set of Profiling-UD metrics --- covering all nine groups in Table~\ref{tab:features} --- is computed for all 
prompts and used in the full predictive analysis of Stage~2.

\subsubsection{Controlled Prompt Generation}

Prompt variants are generated using \textit{Qwen3:32B} with the paraphrasing template shown in Listing~\ref{lst:paraphrase_prompt}, which instructs the model to modify a specific linguistic property while preserving semantic meaning. The choice of \textit{Qwen3:32B} was motivated by two factors. 
First, at the time of the study, Qwen3 ranked among the top open-source instruction-tuned models on established benchmarks~\cite{llmstats2025leaderboard}, making it a strong candidate for following precise linguistic manipulation instructions. 
Second, the 32B parameter size was selected based on preliminary experimentation: smaller models failed to produce consistent variability across surface-level metrics such as token count and sentence count, while the 32B variant reliably responded to targeted paraphrasing instructions within the constraints of our experimental infrastructure\footnote{All experiments were conducted on a machine equipped with an AMD Ryzen 9 7950X 16-core processor, 64 GB of RAM, and an NVIDIA GeForce RTX 4090 GPU (24 GB VRAM)}.
For each of the 30 metrics, three variants are generated per definition pair corresponding to the targets \texttt{minimise}, \texttt{equal}, and \texttt{maximise}, yielding $100 \times 30 \times 3 = 9{,}000$ definition pairs.

To validate the experimental space, we compare the linguistic distributions of the 100 original definition pairs against the 9,000 generated variants. For each metric, we compute descriptive statistics --- mean, standard deviation, interquartile range --- and visualise distributional differences using boxplots. Increased dispersion in the extended dataset indicates that the targeted linguistic properties were successfully manipulated.

\begin{lstlisting}[
caption={Prompt used for controlled linguistic paraphrasing},
label={lst:paraphrase_prompt},
breaklines=true,
basicstyle=\ttfamily\footnotesize
]
Paraphrase the text below. Return only the paraphrased text - no explanations, tags, or comments.

Aim for a paraphrase that reflects the following linguistic feature:

Metric: {metric_name}
Description: {metric_description}
Target: {target_range}
Text: "{functional_def} || {non_functional_def}"

\end{lstlisting}

\subsection{Stage 2: Performance Evaluation and Prediction}


\subsubsection{LLM Evaluation}

The 9,000 prompt variants are evaluated using five open-source instruction-tuned LLMs: \textit{Qwen2-7B-Instruct}, \textit{Falcon3-7B-Instruct}, \textit{Granite-3.2-8B-Instruct}, \textit{Ministral-8B-Instruct-2410}, and \textit{Meta-Llama-3-8B-Instruct}. These models were empirically validated by Zadenoori et al.~\cite{Zadenoori2025}, demonstrating their suitability for LLM-based binary requirements classification and establishing them as a representative benchmark suite for this task.
To measure performance, we focus on F1, F2, precision (P) and recall (R) metrics. These are averaged over the 625 requirements within the dataset~\cite{ClelandHuang2007}, tested over each prompt variant. 

\subsubsection{Regression Modelling and Feature Analysis}

To answer the research questions, we frame prompt performance prediction as a regression problem: each prompt is represented as a feature vector $\mathbf{x} \in \mathbb{R}^{124}$ of Profiling-UD linguistic metrics, and the target $y$ is one of four performance indicators $y \in \{F1, F2, P, R\}$ aggregated over the 625 requirements per prompt variant.

We evaluate five regression models of increasing complexity: \emph{Decision Tree} (DT), \emph{Linear Support Vector Machine} (LSVM), \emph{Random Forest} (RF), \emph{Gradient Boosting} (GB), and \emph{Multi-layer Perceptron} (MLP). This selection covers a spectrum from interpretable linear and rule-based models to non-linear ensembles, following established practice in empirical SE studies that compare multiple model families to avoid bias towards any single algorithm~\cite{HE2015170}. The MLP is wrapped with target standardisation for numerical stability in regression. For each model, hyperparameters are tuned via \emph{Randomised Search} (20 iterations, optimising $R^2$) prior 
to evaluation. 
Model performance is assessed using 10-fold cross-validation stratified by experimental condition (\texttt{metric\_name} $\times$ \texttt{target\_range}), ensuring that prompt variants generated from the same metric and target are distributed across folds proportionally. 

We report three complementary evaluation metrics based on standard regression-based empirical SE studies~\cite{HE2015170, PASCARELLA201922}:

\begin{itemize}
    \item \textbf{Coefficient of determination} ($R^2$): proportion of variance in the target variable explained by the model; ranges from $-\infty$ to 1, where 1 indicates a perfect fit and values below 0 indicate the model performs worse than a constant mean predictor.
    \item \textbf{Mean Absolute Error} (MAE): average absolute difference between predicted and actual values, expressed in the target’s original unit.
    \item \textbf{Root Mean Squared Error} (RMSE): square root of the average squared prediction error; penalises large errors more strongly than MAE, making it sensitive to outliers.
\end{itemize}

To assess statistical significance, we apply a permutation test on $R^2$ with 500 permutations and Benjamini-Hochberg FDR correction across all target-model combinations.


To address \textbf{RQ1}, we evaluate the contribution of each linguistic feature group independently by running the best-performing model using one feature family at a time (as defined in Table~\ref{tab:features}), comparing each subset's $R^2$ against the full-feature baseline. 
This analysis identifies which levels of linguistic description carry the most predictive signal for prompt performance, and whether any single group alone is sufficient to explain a meaningful proportion of variance.

To address \textbf{RQ2}, permutation importance is computed on the best-performing model per target, providing a ranking of the linguistic features that most strongly contribute to each performance indicator. Rankings are computed separately per LLM and then compared across models and targets to identify features with high predictive relevance and to assess the stability of their importance. This analysis allows us to distinguish broadly relevant linguistic predictors from features whose importance varies across models or performance metrics.


\section{Results}
\label{sec:results}

\subsection{Linguistic Variability}

\begin{figure}[h]
    \centering
    \includegraphics[width=\columnwidth]{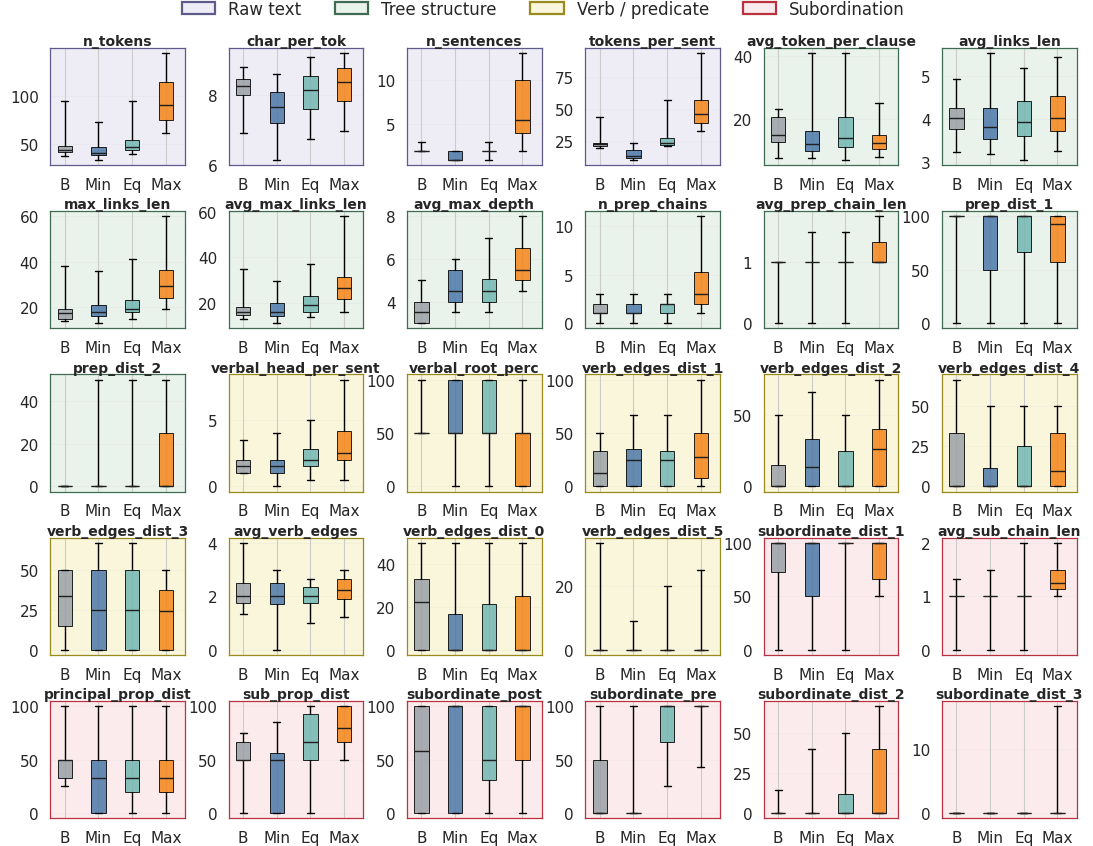}
    \caption{Linguistic variability analysis.}
    \label{fig:linguistic-variability}
\end{figure}

To verify that controlled paraphrasing effectively expanded the experimental space, we compared the baseline prompt definitions (B) against the generated variants under the three control targets \texttt{minimise} (Min), \texttt{equal} (Eq), and \texttt{maximise} (Max) across the 30 targeted metrics identified in Section~\ref{sec:lin-feat-ex}. Results are reported in Figure~\ref{fig:linguistic-variability}.

The distributions show systematic target-dependent shifts for a substantial subset of metrics. 
Quantitatively, 46.2\% of targeted metrics moved in the expected direction under \texttt{minimise} and 61.5\% under \texttt{maximise}. 
Considering stricter bidirectional controllability (i.e., correct movement in both \texttt{minimise} and \texttt{maximise}), 30.8\% of metrics satisfied this criterion. 
In addition, the \texttt{equal} condition was closer to baseline than both extremes for 61.5\% of metrics, indicating partial preservation of intermediate linguistic profiles. 

Partial controllability is explained by structural constraints inherent to the input and the metrics themselves. 
Some metrics are bounded by the nature of the seed definitions. 
For example, since each definition pair consists of one sentence per class, metrics such as \texttt{n\_sentences} have a structural minimum of 1, leaving no meaningful room for \texttt{minimise} to operate. 
Other metrics --- such as \texttt{avg\_prepositional\_chain\_len}, \texttt{prep\_dist\_2}, and \texttt{verb\_edges\_dist\_0} --- already present near-zero baseline values, making \texttt{minimise} ineffective not because the paraphrasing fails, but because the metric has no further room to decrease. 
In both cases, the absence of directional movement reflects a ceiling or floor imposed by the data rather than a failure of the manipulation procedure.

Additionally, the generated variants increased dispersion relative to baseline: the median variability gain was approximately 2.12$\times$ when measured as interquartile range variability (i.e., $\mathrm{IQR}_{\text{controlled}}/\mathrm{IQR}_{\text{baseline}}$), and 88.9\% of the metrics with non-zero baseline IQR showed an increase ($\mathrm{IQR}_{\text{controlled}}/\mathrm{IQR}_{\text{baseline}}>1$). This confirms that controlled paraphrasing generally broadened linguistic coverage, 
providing a diverse experimental space. 

\subsection{Predictive Feasibility (RQ1)}


Table~\ref{tab:rq1-results} reports regression performance across the four performance targets. All model--target combinations are statistically significant ($q < 0.05$, permutation test with Benjamini-Hochberg correction). Random Forest (RF) consistently outperforms all other models, achieving an average of $R^2 = 0.405$ across all target metrics. 
Ensemble methods (RF and GB) substantially outperform linear and shallow models (LSVM, DT), indicating that the relationship between linguistic features and prompt performance is non-linear and benefits from the interaction of multiple features.

\begin{table}[h]
\centering
\caption{Regression model performance across quality targets. 
Best values per target are \textbf{bold}.}
\label{tab:rq1-results}
\small
\setlength{\tabcolsep}{2.4pt}
\begin{tabular}{l l rrr rrr rrr rrr}
\toprule
& & \multicolumn{3}{c}{\textbf{Avg\_F1}} 
& \multicolumn{3}{c}{\textbf{Avg\_F2}} 
& \multicolumn{3}{c}{\textbf{Avg\_Precision}} 
& \multicolumn{3}{c}{\textbf{Avg\_Recall}} \\
\cmidrule(lr){3-5}\cmidrule(lr){6-8}\cmidrule(lr){9-11}\cmidrule(lr){12-14}
& \textbf{Model} 
& $R^2$ & MAE & RMSE 
& $R^2$ & MAE & RMSE 
& $R^2$ & MAE & RMSE 
& $R^2$ & MAE & RMSE \\
\midrule
& DT   
& 0.145 & 0.059 & 0.069 
& 0.141 & 0.060 & 0.070 
& 0.124 & 0.015 & 0.019 
& 0.138 & 0.053 & 0.062 \\
& LSVM 
& 0.164 & 0.059 & 0.069 
& 0.158 & 0.060 & 0.069 
& 0.144 & 0.016 & 0.019 
& 0.155 & 0.053 & 0.061 \\
& MLP  
& 0.290 & 0.051 & 0.063 
& 0.287 & 0.052 & 0.064 
& 0.269 & 0.014 & 0.017 
& 0.284 & 0.046 & 0.056 \\
& GB   
& 0.352 & 0.050 & 0.060 
& 0.350 & 0.051 & 0.061 
& 0.316 & 0.014 & 0.017 
& 0.347 & 0.044 & 0.054 \\
& RF   
& \textbf{0.418} & \textbf{0.045} & \textbf{0.057} 
& \textbf{0.413} & \textbf{0.046} & \textbf{0.058} 
& \textbf{0.377} & \textbf{0.013} & \textbf{0.016} 
& \textbf{0.410} & \textbf{0.041} & \textbf{0.051} \\
\bottomrule
\end{tabular}
\end{table}

Table~\ref{tab:family-contribution} reports, for \textit{Avg\_F1} as a representative performance target, the contribution of each linguistic feature group in isolation compared against the full-feature baseline ($R^2 = 0.418$). Results reveal a two-tier structure. Syntactic Relations ($R^2 = 0.375$) and Morphosyntactic Information ($R^2 = 0.331$) each retain substantial predictive power alone, accounting for most of the signal recoverable from any single group. Parse Tree Structure ($R^2 = 0.210$) contributes moderately. All remaining groups produce near-zero $R^2$. These results suggest that \textbf{syntactic and morphosyntactic features drive predictability}, while surface-level and clausal features might contribute only when combined with the richer feature space.

\begin{table}[t]
\centering
\caption{Feature family contribution to $R^2$ for \textit{Avg\_F1} 
(Random Forest). $\Delta R^2$ is the difference from the full-feature 
baseline. All subsets are statistically significant ($q < 0.05$).}
\label{tab:family-contribution}
\small
\begin{tabular}{l r r r r r}
\toprule
\textbf{Feature subset} & \textbf{\# feat.} & $R^2$ & $\Delta R^2$ & \textbf{MAE} & \textbf{RMSE} \\
\midrule
All features                        & 124 & \textbf{0.418} & --- & 0.045 & 0.057 \\
\midrule
Syntactic Relations                 & 37  & 0.375 & $-$0.044 & 0.047 & 0.059 \\
Morphosyntactic Information         & 33  & 0.331 & $-$0.087 & 0.049 & 0.061 \\
Global \& Local Parse Tree Struct.  &  9  & 0.210 & $-$0.208 & 0.055 & 0.067 \\
Inflectional Morphology             & 14  & 0.024 & $-$0.394 & 0.063 & 0.074 \\
Order of Elements                   &  4  & 0.004 & $-$0.415 & 0.065 & 0.075 \\
Raw Text Properties                 &  4  & 0.003 & $-$0.415 & 0.061 & 0.075 \\
Lexical Variety                     &  6  & 0.001 & $-$0.417 & 0.065 & 0.075 \\
Use of Subordination                &  8  & 0.000 & $-$0.418 & 0.065 & 0.075 \\
Verbal Predicate Structure          &  9  & $-$0.005 & $-$0.423 & 0.064 & 0.075 \\
\bottomrule
\end{tabular}
\end{table}

\begin{rqbox}[Answer to RQ1]{cyan}
\begin{rqlist}
\item Linguistic features significantly predict prompt performance across all targets ($q < 0.05$, $R^2 \in [0.38, 0.42]$ for RF), providing empirical support for the main hypothesis.
\item Predictive power requires the full feature space. Syntactic Relations and Morphosyntactic Information contribute most individually, suggesting that sentence structure and part-of-speech 
diversity are the primary linguistic drivers of prompt performance.
\item An $R^2 \approx 0.41$ is particularly relevant given that many other factors influence LLM output beyond linguistic structure. Linguistic profiling combined with standard regression models provides an interpretable, LLM-free signal for prompt characterisation, complementary to empirical LLM evaluation.
\end{rqlist}
\end{rqbox}

\subsection{Linguistic Predictor Importance (RQ2)}

\begin{figure}[h]
    \centering
    \begin{subfigure}[t]{\columnwidth}
        \centering
        \includegraphics[width=\columnwidth]{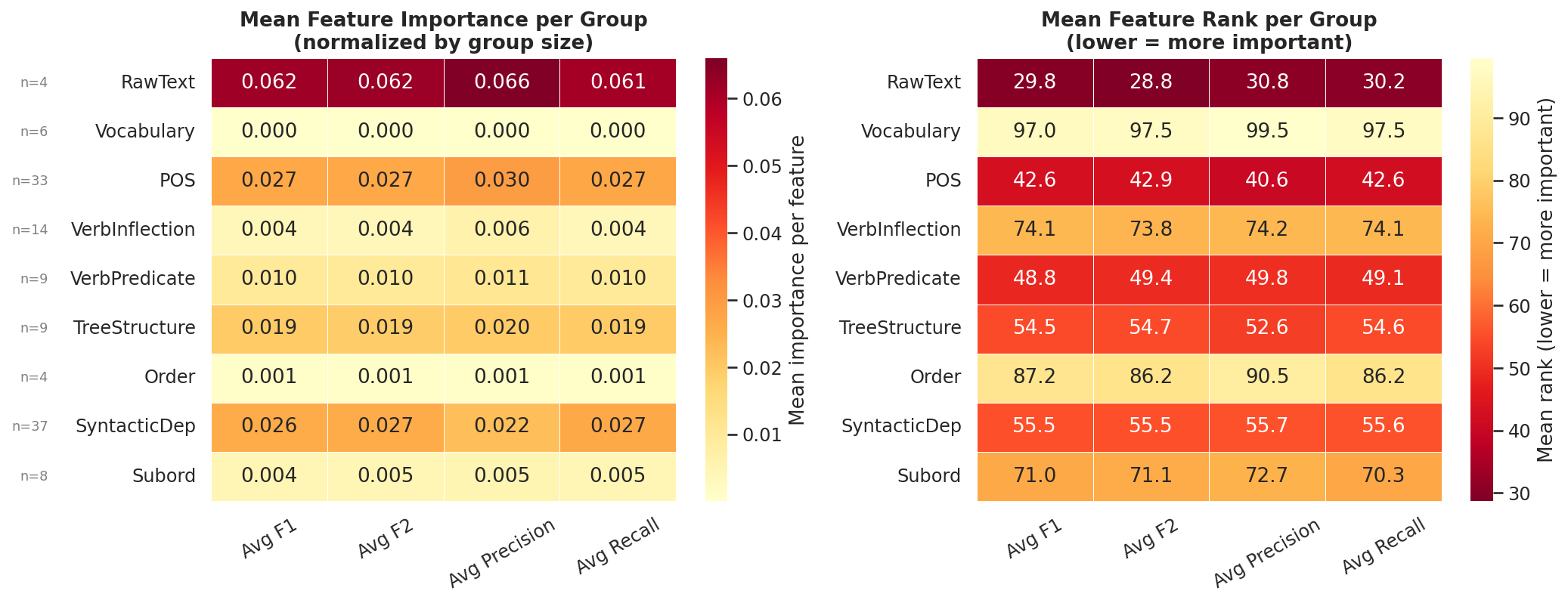}
        \caption{Importance and rank analysis.}
        \label{fig:group-importance}
    \end{subfigure}
    \vspace{4pt}
    \begin{subfigure}[t]{\columnwidth}
        \centering
        \includegraphics[width=\columnwidth]{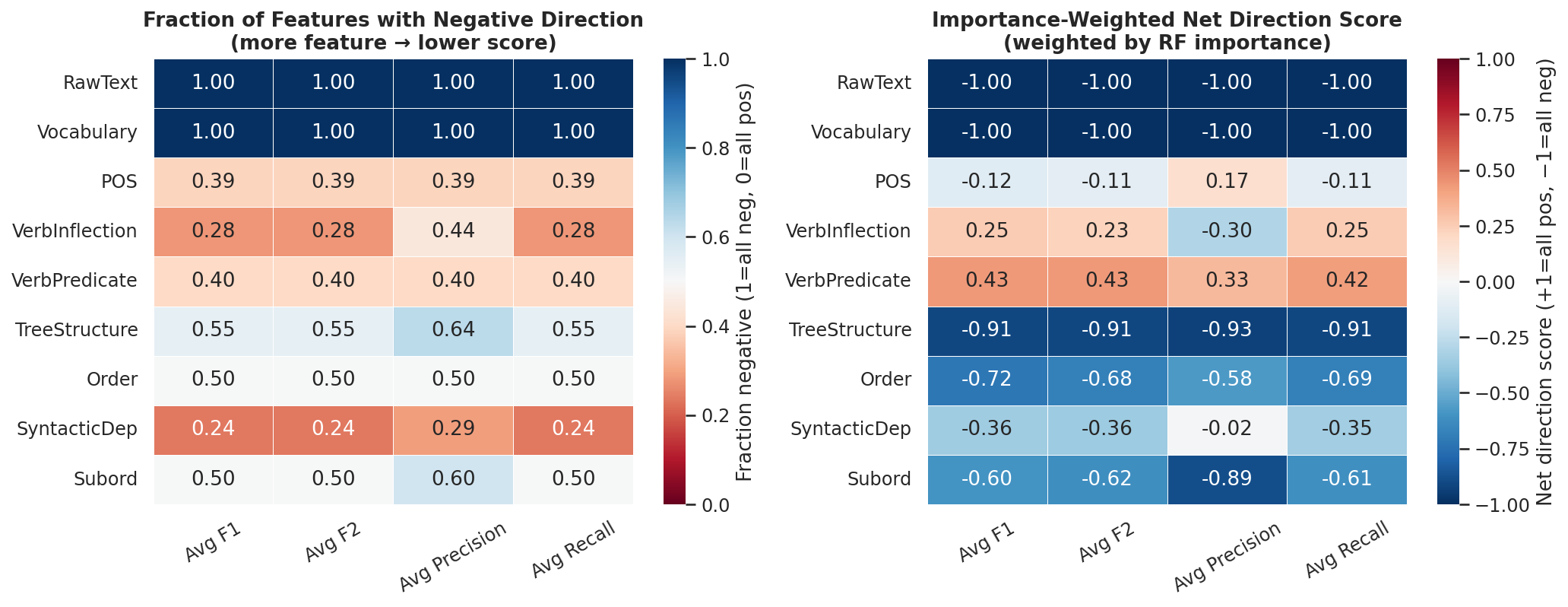}
        \caption{Direction analysis.}
        \label{fig:group-direction}
    \end{subfigure}
    \caption{Linguistic feature group analysis.}
    \label{fig:group-analysis}
\end{figure}

Figure~\ref{fig:group-analysis} reports group-level feature importance and direction profiles across the four performance targets. Raw Text features achieve the highest per-feature importance despite comprising only four features (mean importance $\approx 0.062$), while Syntactic Dependencies and POS features contribute most in absolute terms. Vocabulary features (TTR variants) are irrelevant across all targets (mean importance $\approx 0.000$, mean rank $> 95$). The direction profile confirms that Raw Text is fully and consistently negative (weighted score $= -1.0$): longer definitions with denser words and more sentences systematically reduce performance. Tree Structure features are similarly negative ($-0.91$), while POS shows mixed directionality.

\begin{figure}[h] 
    \centering 
        \includegraphics[width=0.8\columnwidth]{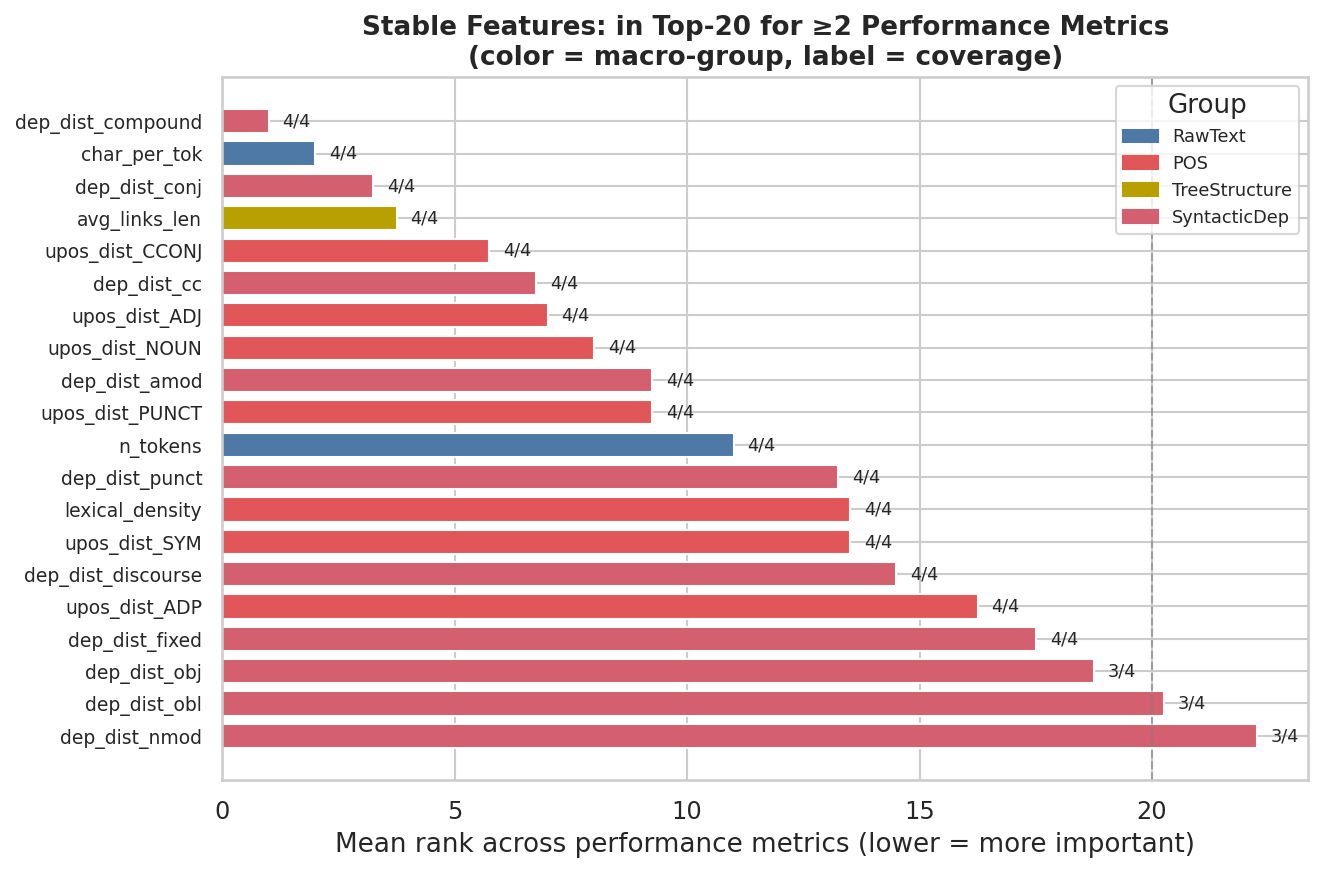} \caption{Linguistic feature analysis on top-20 stable features.} 
        \label{fig:stable-top} 
\end{figure}

Figure~\ref{fig:stable-top} shows the top-20 features stable across all four performance targets. These come exclusively from Syntactic Dependencies, POS, Tree Structure and Raw Text features. The top-ranked feature is compound dependency distribution (\texttt{dep\_dist\_compound}, mean rank 1.0, importance $0.21$--$0.33$) --- the proportion of dependencies forming noun compounds.
Character-per-token ratio (\texttt{char\_per\_tok}, rank 2.0) and density of conjunction 
(\texttt{dep\_dist\_conj}, \texttt{upos\_dist\_CCONJ}, ranks 3.2--5.8) follow.

Qualitative inspection reveals that these features trace domain-specific technical language: the most frequent compound pairs include \textit{quality $\to$ requirements}, \textit{performance $\to$ efficiency}, and \textit{system $\to$ operations}, while the dominant long tokens are ISO/IEC~25010 quality attribute names (\textit{maintainability, compatibility, portability}).
The dominant coordinating conjunction is \textit{and}, coordinating nominal chains of up to 23 items.
These prompts are technically dense, noun-compound-heavy, and enumerative, which reflects the nature of RE documentation.

\begin{figure}[h] 
    \centering 
        \includegraphics[width=0.750\columnwidth]{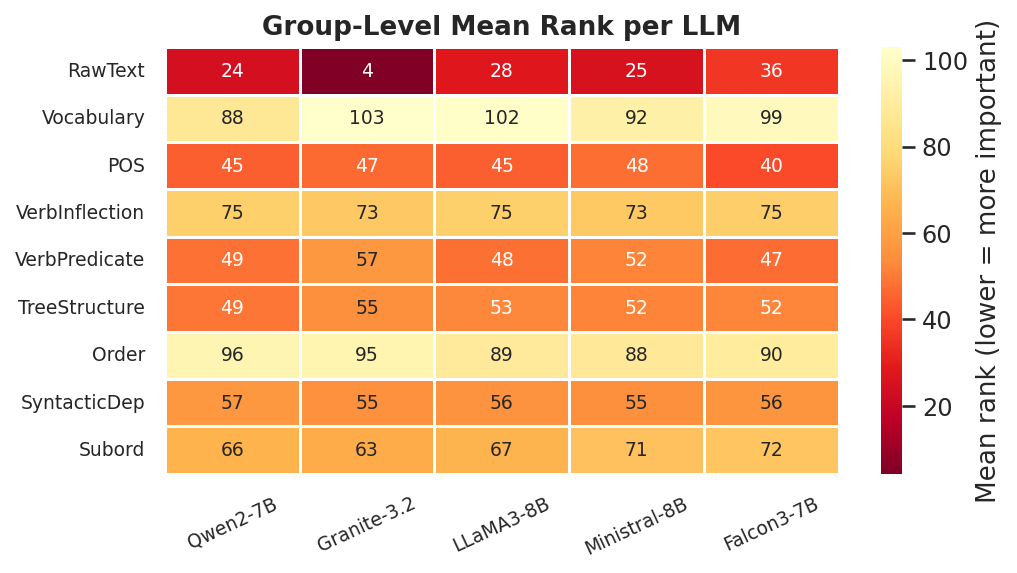} \caption{Group-level, cross-LLM mean rank analysis.} 
        \label{fig:group-llm} 
\end{figure}
\begin{figure}[h] 
    \centering 
        \includegraphics[width=0.83\columnwidth]{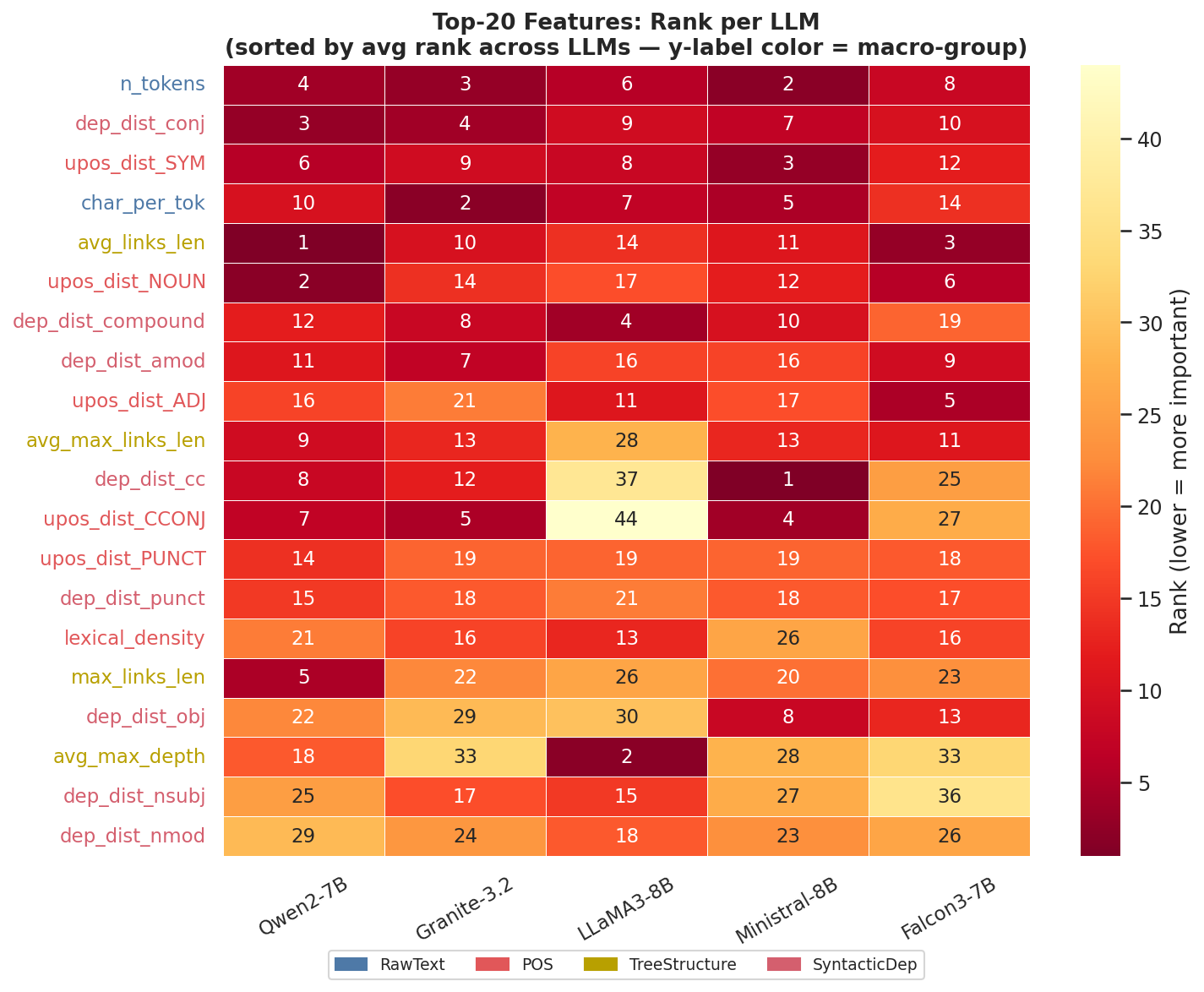} \caption{Ranked cross-LLM analysis of top linguistic features.} 
        \label{fig:heatmap} 
\end{figure}

Figures~\ref{fig:group-llm} and~\ref{fig:heatmap} report cross-LLM consistency. Pairwise Spearman $\rho$ between LLM-specific feature rankings ranges from 0.86 to 0.95, and rank consistency across the four performance targets reaches $\rho \geq 0.97$, confirming that the identified predictors are both metric-agnostic and largely model-agnostic. The most universally stable individual predictors are \texttt{char\_per\_tok}, \texttt{dep\_dist\_conj}, and \texttt{avg\_links\_len}, which appear in the top-10 for all five LLMs. Within this overall consensus, two patterns of model-specific divergence are worth noting. First, token count (\texttt{n\_tokens}) ranks \#1 for \textit{Meta-Llama-3-8B-Instruct} but falls to ranks 12--51 for the remaining models, suggesting that LLaMA-3 is disproportionately sensitive to absolute prompt length --- possibly reflecting differences in instruction-tuning data composition or in how few-shot context is handled during training. Second, pronoun and determiner distributions show rank standard deviations above 20 across LLMs, indicating that the way function-word categories are weighted varies substantially from model to model, even when the broader structural signal is stable. These cases indicate that, while the dominant linguistic predictors are model-agnostic, fine-grained predictor importance still carries useful information for model-specific prompt tuning.

\begin{rqbox}[Answer to RQ2]{cyan}
\begin{rqlist}
\item Feature importance rankings are highly consistent across all five LLMs (pairwise Spearman $\rho \in [0.86, 0.95]$) and across all four performance targets ($\rho \geq 0.97$).
\item The most consistently harmful features are compound dependency distribution (\texttt{dep\_dist\_compound}, rank \#1), character-per-token ratio (\texttt{char\_per\_tok}, rank \#2), and conjunction density (\texttt{dep\_dist\_conj}, \texttt{upos\_dist\_CCONJ}). 
\item Lexical diversity features (TTR variants) are universally irrelevant (mean rank $>95$ for all models and targets), and verbal inflection and subordination features are similarly uninformative. Vocabulary variability does not predict prompt performance.
\item While cross-LLM consensus is high, some features show model-specific importance. This confirms that LLM selection remains a relevant design variable. 
\end{rqlist}
\end{rqbox}

\section{Discussion}
\label{sec:discussion}

The results of RQ1 and RQ2 jointly show that prompt performance is not only affected by the semantic content of a prompt, but also by how that content is linguistically packaged. 
RQ1 shows that linguistic features provide a significant predictive signal for performance, while 
RQ2 shows that this signal is concentrated in a specific structural profile --- morphosyntactic patterns, long dependency arcs and dense coordination --- rather than being uniformly distributed across linguistic dimensions. Notably, these are some of the same structural properties associated with increased processing cost in human language comprehension, from dependency-length effects \cite{futrell2015large} to the nominalisation patterns characteristic of expository and technical writing \cite{halliday1994writing}. This convergence reinforces the interpretability dimension that motivated our approach: the features driving prompt performance are not opaque statistical artefacts of a particular model, but linguistic properties with independent theoretical grounding. \textbf{Prompts should therefore be treated as SE artefacts that humans can analyse, reason about, and refine using well-established linguistic categories}, a perspective that complements automatic optimisation. Moreover, these observations connect to a recent line of work on LLM linguistic fragility: Cao et al. \cite{cao-etal-2026-style} showed that surface-level stylistic variation (formality, readability, politeness) degrades RAG performance through retrieval and generation errors. Our results extend this picture along a complementary axis --- the fragility is not only stylistic but structural, and it can be characterised through measurable syntactic and morphosyntactic properties before any LLM is invoked. This, in turn, motivates a different role for linguistic profiling in prompt engineering: not as a post-hoc diagnostic, but as a pre-inference scoring signal.

Raw Text features have high per-feature importance (Figure~\ref{fig:group-analysis}) yet weak standalone predictive power (Table~\ref{tab:family-contribution}). 
\textbf{Length, sentence count, and word length therefore operate as proxies for deeper structural density rather than as explanatory factors in themselves}, drawing their predictive weight from associated phenomena --- compound noun phrases, long dependency arcs, coordinated nominal lists, and dense technical terminology. 
This pattern reflects the register of technical documentation: in our study, the patterns are tied to requirements and quality-attribute terminology, but the implication is broader. 
\textbf{Prompts including content copied from standards, taxonomies, or expert documentation may be conceptually accurate while being linguistically difficult for LLMs to process}, creating a tension between domain completeness and prompt usability. Making definitions more precise, exhaustive, and terminology-rich can simultaneously introduce the structural patterns that reduce prompt effectiveness. 
\textbf{Expert knowledge should therefore be adapted rather than used verbatim}: the prompt should preserve decision-relevant cues while reducing unnecessary syntactic and nominal density --- but not unnecessary lexical variability. RQ2 shows that lexical-variety features (TTR variants) are universally irrelevant predictors, indicating that vocabulary diversity, often promoted as a writing virtue, carries no measurable benefit for prompt effectiveness.

These findings yield actionable implications for prompt engineering in SE:

\begin{enumerate}
    \item \textbf{Prefer simple over exhaustive definitions.}
    Prompt definitions should support the target decision rather than reproduce the full conceptual richness of the domain.

    \item \textbf{Express one idea at a time.}
    Avoid compressing several criteria, exceptions, or distinctions into a single long sentence.

    \item \textbf{Avoid long additive enumerations.}
    Long lists joined by additive connectors such as \textit{and} should be shortened, grouped, or split into separate statements.

    \item \textbf{Reduce compound-heavy noun phrases.}
    Unpack compressed technical expressions when doing so improves readability, e.g., ``requirements about system quality'' instead of ``system quality requirements''.

    \item \textbf{Use domain-specific terminology selectively.}
    Keep technical terms that are necessary for the task, but avoid terminology that only increases domain density.

    \item \textbf{Do not optimise prompt length in isolation.}
    Length is a warning signal, but the main risks are the structures that often accompany it: compounds, coordination chains, long dependency arcs, and dense nominal constructions.

    \item \textbf{Do not optimise for lexical variety.}
    Replacing words merely to increase vocabulary richness is unlikely to help; consistency is preferable to lexical embellishment.

    \item \textbf{Penalise dense syntactic structures.}
    Candidate prompts with high compound density, long dependency arcs, high coordination density, or dense nominal constructions can be deprioritised before LLM evaluation.
\end{enumerate}










\section{Threats to validity}
\label{sec:ttv}

We organise threats to validity following the taxonomy by Wohlin et al.~\cite{Wohlin2012}.

\emph{Internal validity.} Linguistic variability is generated exclusively using \textit{Qwen3:32B}, without comparing alternative paraphrasing models. Prompt generation is instrumental to the study, not its object of investigation. We empirically validate that the generated variants produce significant expansion across multiple metrics, which is enough for our purposes. Future work may assess different generation models and how they affect linguistic distributions.

\emph{Construct validity.} We operationalise linguistic predictive importance through permutation importance. This measure is sensitive to feature correlation: when linguistic features are correlated, importance may be distributed arbitrarily between them, potentially underestimating the contribution of individual features. 
We partially mitigate this by reporting group-level contributions alongside individual rankings. 
Additionally, we operationalise prompt performance through four performance indicators derived from a binary classification task, which might not generalise to measure prompt quality in other contexts. 
A further construct threat concerns potential data leakage in the PROMISE-NFR dataset~\cite{PROMISE}, since some requirements may have appeared in the pre-training data of the evaluated LLMs. 
To mitigate this, we conducted an additional leakage check by prompting each LLM with the first half of each requirement and comparing its generated continuation with the original second half using Jaccard similarity. 
Mean Jaccard scores remain low across models, ranging within the interval $[0.105, 0.165]$, suggesting no strong evidence of systematic reconstruction of hidden requirement segments. 
Full results are available in the replication package.

\emph{External validity.} The study is scoped to a single task (binary requirements classification), dataset (PROMISE-NFR~\cite{PROMISE}), and prompt component (class definitions). This is a conscious design choice: rather than varying these dimensions simultaneously --- which would reduce experimental control --- we conduct an in-depth analysis across 9,000 prompt variants, 124 linguistic metrics, five LLMs and five regression models. 
The statistical significance and consistent findings across these dimensions validate our hypothesis in a well-controlled context, establishing a replicable pipeline. Generalisation to other tasks, datasets and prompting strategies remains an open question that this work is explicitly designed to motivate. 

\emph{Conclusion validity.} The five LLMs used in this study were selected based on their empirical validation for requirements classification~\cite{Zadenoori2025}. While newer models may outperform absolute metric values, we observe consistent patterns across architecturally diverse models. This suggests that the identified linguistic predictors reflect properties of the task and prompt structure rather than idiosyncrasies of specific models. 

\section{Related work}
\label{sec:rel-work}

Prompt engineering is now a practical requirement for applying LLMs to SE tasks. Recent studies examine prompting for code generation~\cite{khojah2025impact,bruni2025benchmarking}, unit test generation~\cite{ouedraogo2026prompt}, repository mining~\cite{de2025framework}, and software startup support~\cite{ahlgren2025assisting}. In RE, prompt engineering has been studied across tasks such as elicitation, validation, traceability, and classification~\cite{huang2025prompt}. Binkhonain and Alfayaz show that prompt-based LLMs, especially few-shot and persona-based variants, can perform competitively with fine-tuned models while reducing the need for task-specific annotated data~\cite{binkhonain2025prompts}. 
These studies establish that prompt design affects SE outcomes and that prompt engineering is a viable alternative to retraining in data-scarce settings.

However, existing SE research mostly treats prompts as task templates or strategy configurations. The usual unit of analysis is the prompting technique: zero-shot, few-shot, persona, chain-of-thought, task-specific prompting, or their combinations~\cite{khojah2025impact,binkhonain2025prompts}. Other studies focus on end-to-end workflows, such as prompt refinement for repository mining~\cite{de2025framework}, prompt patterns for startup support~\cite{ahlgren2025assisting}, or benchmark comparisons for test and code generation~\cite{ouedraogo2026prompt,bruni2025benchmarking}. This literature shows that prompt structure influences task performance, but it does not explain which linguistic properties of prompts matter. In particular, SE studies rarely analyse prompts in terms of syntactic structure, morphosyntactic composition, dependency patterns, lexical density, or surface complexity. 

Work in NLP has started to address this linguistic dimension. Leidinger et al. analyse semantically equivalent prompts varying in mood, tense, aspect, modality, and synonym choice, showing that prompt performance varies across models and datasets and is not reliably explained by prompt length, perplexity, word frequency, or ambiguity~\cite{leidinger2023languagepromptinglinguisticproperties}. Wahle et al. study controlled paraphrase types across morphology, syntax, lexicon, lexico-syntax, discourse, and other categories, showing that linguistic reformulations can change model behaviour across tasks and models~\cite{wahle-etal-2024-paraphrase}. Long et al. propose a broader property-centric view of prompt quality, including linguistic and human-oriented properties beyond output accuracy~\cite{long-etal-2025-makes}. Cao et al. show that linguistic style variations such as formality, readability, politeness, and grammatical correctness can degrade RAG performance through retrieval and generation error propagation~\cite{cao-etal-2026-style}. These studies show that linguistic form affects LLM behaviour, but they mostly analyse coarse linguistic categories or style dimensions. They also diagnose sensitivity after inference rather than using linguistic features to predict prompt effectiveness or guide APE search beforehand.

This paper addresses these gaps by treating prompts as measurable linguistic artefacts. We link prompt linguistic features to LLM performance for SE tasks, identifying predictors useful for prompt selection, rewriting, and APE pre-filtering before costly inference.

\section{Conclusions}
\label{sec:conclusions}


Our findings suggest an alternative to computationally expensive, opaque APE approaches: linguistic profiling can serve as a fast, human-interpretable pre-filter that scores candidate prompts before any LLM is invoked. Stable linguistic predictors can be computed from raw text in milliseconds using off-the-shelf tools. An APE pipeline that incorporates this prior could deprioritise structurally dense candidates early, reducing the number of LLM evaluations required to reach a high-performing prompt. This is particularly relevant in resource-constrained or energy-sensitive deployment contexts.

Several directions follow from this work. 
First, the experimental scope should be expanded along three axes: replication across other RE and SE tasks; extension of linguistic manipulation to the full prompt structure, covering system context, task instruction, and output specification; and evaluation on closed-source and larger-scale LLMs.
Second, our framework manipulates prompts along independent linguistic dimensions; future work could explore joint manipulation of multiple features and study interaction effects, which our regression results suggest are non-trivial.
Third, the catalogue of linguistic predictors could be integrated into APE toolchains as a scoring function, enabling hybrid approaches combining linguistic screening with semantic search over candidate prompts.
Finally, the findings motivate the development of linguistically-aware prompt rewriting tools for SE practitioners.



\section*{Data Availability}

All data, prompts, model outputs, scripts, results and instructions needed to reproduce the study are available in an anonymised replication package: \url{https://anonymous.4open.science/r/ESEM-D64C/}. 
The package includes a README describing the processing pipeline and how to reproduce the analyses.





\bibliography{refs}

@inproceedings{miaschi-etal-2020-linguistic,
    title = "Linguistic Profiling of a Neural Language Model",
    author = "Miaschi, Alessio  and
      Brunato, Dominique  and
      Dell{'}Orletta, Felice  and
      Venturi, Giulia",
    booktitle = "28th International Conference on Computational Linguistics",
    year = "2020"
}

@inproceedings{miaschi-etal-2024-evaluating,
    title = "Evaluating Large Language Models via Linguistic Profiling",
    author = "Miaschi, Alessio  and
      Dell{'}Orletta, Felice  and
      Venturi, Giulia",
    booktitle = "Empirical Methods in Natural Language Processing",
    year = "2024"
}

@inproceedings{marvin2023prompt,
  title={Prompt engineering in large language models},
  author={Marvin, Ggaliwango and Hellen, Nakayiza and Jjingo, Daudi and Nakatumba-Nabende, Joyce},
  booktitle={International conference on data intelligence and cognitive informatics},
  year={2023},
  pages={387--402}
}

@article{ouedraogo2026prompt,
  title={{Prompt engineering in LLMs for automated unit test generation: A large-scale study}},
  author={Ou{\'e}draogo, Wendk{\^u}uni and Kabor{\'e}, Abdoul and Li, Yinghua and Tian, Haoye and Koyuncu, Anil and Klein, Jacques and Lo, David and Bissyand{\'e}, Tegawend{\'e} F},
  journal={Empirical Software Engineering},
  volume={31},
  number={4},
  year={2026},
  pages={103}
}

@article{khojah2025impact,
  title={The impact of prompt programming on function-level code generation},
  author={Khojah, Ranim and de Oliveira Neto, Francisco Gomes and Mohamad, Mazen and Leitner, Philipp},
  journal={IEEE Transactions on Software Engineering},
  year={2025}
}

@article{taherkhani2024automated,
  title={Automated Prompt Engineering for Cost-Effective Code Generation Using Evolutionary Algorithms},
  author={Taherkhani, Hamed and Sepidband, Melika and Pham, Hung Viet and Wang, Song and Hemmati, Hadi},
  journal={ACM Transactions on Software Engineering and Methodology},
  year={2024}
}

@inproceedings{bruni2025benchmarking,
  title={Benchmarking prompt engineering techniques for secure code generation with gpt models},
  author={Bruni, Marc and Gabrielli, Fabio and Ghafari, Mohammad and Kropp, Martin},
  booktitle={2025 IEEE/ACM Second International Conference on AI Foundation Models and Software Engineering (Forge)},
  year={2025},
  pages={93--103}
}

@article{binkhonain2025prompts,
  title={{Are prompts all you need? Evaluating prompt-based Large Language Models (LLM) s for software requirements classification}},
  author={Binkhonain, Manal and Alfayez, Reem},
  journal={Requirements Engineering},
  volume={30},
  number={4},
  year={2025},
  pages={423--443}
}

@article{ahlgren2025assisting,
  title={{Assisting early-stage software startups with LLMs: Effective prompt engineering and system instruction design}},
  author={Ahlgren, Thea Lovise and Sunde, Helene F{\o}nstelien and Kemell, Kai-Kristian and Nguyen-Duc, Anh},
  journal={Information and Software Technology},
  year={2025},
  pages={107832}
}

@inproceedings{shin2025prompt,
  title={Prompt engineering or fine-tuning: An empirical assessment of llms for code},
  author={Shin, Jiho and Tang, Clark and Mohati, Tahmineh and Nayebi, Maleknaz and Wang, Song and Hemmati, Hadi},
  booktitle={2025 IEEE/ACM 22nd International Conference on Mining Software Repositories (MSR)},
  year={2025},
  pages={490--502}
}

@inproceedings{huang2025prompt,
  title={Prompt engineering for requirements engineering: A literature review and roadmap},
  author={Huang, Kaicheng and Wang, Fanyu and Huang, Yutan and Arora, Chetan},
  booktitle={2025 IEEE 33rd International Requirements Engineering Conference Workshops (REW)},
  year={2025},
  pages={548--557}
}

@inproceedings{de2025framework,
  title={A framework for using llms for repository mining studies in empirical software engineering},
  author={De Martino, Vincenzo and Castano, Joel and Palomba, Fabio and Franch, Xavier and Mart{\'\i}nez-Fern{\'a}ndez, Silverio},
  booktitle={2025 IEEE/ACM International Workshop on Methodological Issues with Empirical Studies in Software Engineering (WSESE)},
  year={2025},
  pages={6--11}
}

@article{de2021universal,
  title={Universal dependencies},
  author={De Marneffe, Marie-Catherine and Manning, Christopher D and Nivre, Joakim and Zeman, Daniel},
  journal={Computational linguistics},
  volume={47},
  number={2},
  year={2021},
  pages={255--308}
}

@inproceedings{brunato2020profiling,
  title={{Profiling-UD: A Tool for Linguistic Profiling of Texts}},
  author={Brunato, Dominique and Cimino, Andrea and Dell’Orletta, Felice and Venturi, Giulia and Montemagni, Simonetta},
  booktitle={Twelfth Language Resources and Evaluation Conference},
  year={2020},
  pages={7145--7151}
}

@article{stol2018abc,
  title={{The ABC of software engineering research}},
  author={Stol, Klaas-Jan and Fitzgerald, Brian},
  journal={ACM Transactions on Software Engineering and Methodology},
  volume={27},
  number={3},
  year={2018},
  pages={1--51}
}

@InProceedings{Zadenoori2025,
author="Zadenoori, Mohammad Amin
and Zhao, Liping
and Alhoshan, Waad
and Ferrari, Alessio",
title="Automatic Prompt Engineering: The Case of Requirements Classification",
booktitle="Requirements Engineering: Foundation for Software Quality",
year="2025",
isbn="978-3-031-88531-0",
pages="217--225",
}

@techreport{iso25010,
  type        = {Standard},
  author      = {{ISO/IEC JTC 1/SC 7}},
  key         = {ISO/IEC 25010:2023},
  year        = {2023},
  title       = {{Systems and software 
                  Quality Requirements and Evaluation (SQuaRE) --- Product 
                  quality model}},
  volume      = {2023},
  institution = {International Organization for Standardization}
}

@article{Hou2024,
author = {Hou, Xinyi and Zhao, Yanjie and Liu, Yue and Yang, Zhou and Wang, Kailong and Li, Li and Luo, Xiapu and Lo, David and Grundy, John and Wang, Haoyu},
title = {Large Language Models for Software Engineering: A Systematic Literature Review},
year = {2024},
issue_date = {November 2024},
volume = {33},
number = {8},
issn = {1049-331X},
journal = {ACM Trans. Softw. Eng. Methodol.},
articleno = {220}
}

@inproceedings{errica-etal-2025-wrong,
    title = "What Did {I} Do Wrong? Quantifying {LLM}s' Sensitivity and Consistency to Prompt Engineering",
    author = "Errica, Federico  and
      Sanvito, Davide  and
      Siracusano, Giuseppe  and
      Bifulco, Roberto",
    booktitle = "2025 Conference of the Nations of the Americas Chapter of the Association for Computational Linguistics: Human Language Technologies",
    year = "2025",
    ISBN = "979-8-89176-189-6"
}

@article{mizrahi-etal-2024-state,
    title = {{State of What Art? A Call for Multi-Prompt {LLM} Evaluation}},
    author = "Mizrahi, Moran  and
      Kaplan, Guy  and
      Malkin, Dan  and
      Dror, Rotem  and
      Shahaf, Dafna  and
      Stanovsky, Gabriel",
    journal = "Transactions of the Association for Computational Linguistics",
    volume = "12",
    year = "2024"
}

@misc{sclar2024quantifyinglanguagemodelssensitivity,
      title={{Quantifying Language Models' Sensitivity to Spurious Features in Prompt Design or: How I learned to start worrying about prompt formatting}}, 
      author={Melanie Sclar and Yejin Choi and Yulia Tsvetkov and Alane Suhr},
      year={2024},
      eprint={2310.11324},
      archivePrefix={arXiv},
      primaryClass={cs.CL}
}

@Inbook{Ronanki2024,
author="Ronanki, Krishna
and Cabrero-Daniel, Beatriz
and Horkoff, Jennifer
and Berger, Christian",
title="Requirements Engineering Using Generative AI: Prompts and Prompting Patterns",
bookTitle="Generative AI for Effective Software Development",
year="2024",
isbn="978-3-031-55642-5",
}

@article{Wang2025,
author = {Wang, Guoqing and Sun, Zeyu and Ye, Sixiang and Gong, Zhihao and Chen, Yizhou and Zhao, Yifan and Liang, Qingyuan and Hao, Dan},
title = {Do advanced language models eliminate the need for prompt engineering in software engineering?},
year = {2025},
issn = {1049-331X},
journal = {ACM Trans. Softw. Eng. Methodol.}
}

@inproceedings{Ronanki2025,
author = {Ronanki, Krishna and Arvidsson, Simon and Axell, Johan},
title = {Prompt Engineering Guidelines for Using Large Language Models in Requirements Engineering},
year = {2025},
isbn = {978-3-032-04206-4},
booktitle = {Software Engineering and Advanced Applications: 51st Euromicro Conference, SEAA 2025},
pages={245–262}
}

@inproceedings{ye-etal-2024-prompt,
    title = "Prompt Engineering a Prompt Engineer",
    author = "Ye, Qinyuan  and
      Ahmed, Mohamed  and
      Pryzant, Reid  and
      Khani, Fereshte",
    booktitle = "Findings of the ACL",
    year = "2024",
    pages={490-502}
}

@inproceedings{Liu2024,
author = {Liu, Dairui and Yang, Boming and Du, Honghui and Greene, Derek and Hurley, Neil and Lawlor, Aonghus and Dong, Ruihai and Li, Irene},
title = {RecPrompt: A Self-tuning Prompting Framework for News Recommendation Using Large Language Models},
year = {2024},
booktitle = {33rd ACM International Conference on Information and Knowledge Management},
pages={3902–3906}
}

@inproceedings{kong-etal-2024-prewrite,
    title = "{PR}ewrite: Prompt Rewriting with Reinforcement Learning",
    author = "Kong, Weize  and
      Hombaiah, Spurthi  and
      Zhang, Mingyang  and
      Mei, Qiaozhu  and
      Bendersky, Michael",
    booktitle = "62nd Annual Meeting of the Association for Computational Linguistics (Volume 2: Short Papers)",
    year = "2024"
}

@misc{zhou2023largelanguagemodelshumanlevel,
      title={Large Language Models Are Human-Level Prompt Engineers}, 
      author={Yongchao Zhou and Andrei Ioan Muresanu and Ziwen Han and Keiran Paster and Silviu Pitis and Harris Chan and Jimmy Ba},
      year={2023},
      eprint={2211.01910},
      archivePrefix={arXiv},
      primaryClass={cs.LG}
}

@inproceedings{Rubei2025,
author = {Rubei, Riccardo and Moussaid, Aicha and Di Sipio, Claudio and Di Ruscio, Davide},
title = {Prompt engineering and its implications on the energy consumption of Large Language Models},
year = {2025},
booktitle = {IEEE/ACM 9th International Workshop on Green and Sustainable Software}

,pages={60–67}
}

@INPROCEEDINGS {Cui2025,
author = { Cui, Xing and Wu, Jingzheng and Ling, Xiang and Luo, Tianyue },
booktitle = { 2025 ACM/IEEE International Symposium on Empirical Software Engineering and Measurement},
title = {{ We Know What You're Looking For: Recommendation for Large-Scale Open Source Software }},
year = {2025},
volume = {},
ISSN = {}}

@INPROCEEDINGS {Salman2025,
author = { Salman, Iflaah and Waseem, Muhammad and Mandic, Vladimir and De Alwis, Rasanjana Dhanushkha },
booktitle = { 2025 ACM/IEEE International Symposium on Empirical Software Engineering and Measurement},
title = {{ A Vision for Debiasing Confirmation Bias in Software Testing via LLM }},
year = {2025},
volume = {},
ISSN = {}}

@INPROCEEDINGS{Rahman2025,
  author={Rahman, Md Nafiu and Ahmed, Sadif and Wahab, Zahin and Sohan, S M and Shahriyar, Rifat},
  booktitle={2025 ACM/IEEE International Symposium on Empirical Software Engineering and Measurement}, 
  title={{Secret Breach Detection in Source Code with Large Language Models}}, 
  year={2025},
  volume={},
  number={},
  pages={207-217}
}

@inproceedings{Felizardo2024,
author = {Felizardo, Katia R. and Lima, M\'{a}rcia S. and Deizepe, Anderson and Conte, Tayana U. and Steinmacher, Igor},
title = {{ChatGPT application in Systematic Literature Reviews in Software Engineering: an evaluation of its accuracy to support the selection activity}},
year = {2024},
booktitle = {18th ACM/IEEE International Symposium on Empirical Software Engineering and Measurement},

  pages={25–36}
}

@misc{llmstats2025leaderboard,
  author       = {{LLM Stats}},
  title        = {{Open LLM Leaderboard}},
  year         = {2025},
  howpublished = {\url{https://llm-stats.com/leaderboards/open-llm-leaderboard}},
  note         = {Accessed: Jan 2025}
}

@article{HE2015170,
title = {An empirical study on software defect prediction with a simplified metric set},
journal = {Inf. and Soft. Tech.},
volume = {59},
year = {2015},
issn = {0950-5849},
author = {Peng He and Bing Li and Xiao Liu and Jun Chen and Yutao Ma}

,pages={170-190}
}

@article{PASCARELLA201922,
title = {Fine-grained just-in-time defect prediction},
journal = {Journal of Systems and Software},
volume = {150},
year = {2019},
issn = {0164-1212},
author = {Luca Pascarella and Fabio Palomba and Alberto Bacchelli}

,pages={22-36}
}

@article{Zhao2021,
author = {Zhao, Liping and Alhoshan, Waad and Ferrari, Alessio and Letsholo, Keletso and Ajagbe, Muideen and Chioasca, Erol-Valeriu and Batista-Navarro, Riza},
title = {{Natural Language Processing for Requirements Engineering: A Systematic Mapping Study}},
year = {2021},
volume = {54},
number = {3},
issn = {0360-0300},
journal = {ACM Comput. Surv.},
articleno = {55}
}

@misc{Motger2026,
      title={{Characterizing Datasets for LLM-based Requirements Engineering: A Systematic Mapping Study}}, 
      author={Quim Motger and Carlota Catot and Xavier Franch},
      year={2026},
      eprint={2510.18787},
      archivePrefix={arXiv},
      primaryClass={cs.SE}
}

@book{Wohlin2012,
author = {Wohlin, Claes and Runeson, Per and Hst, Martin and Ohlsson, Magnus C. and Regnell, Bjrn and Wessln, Anders},
title = {Experimentation in Software Engineering},
year = {2012},
isbn = {3642290434}}

@misc{zadenoori2025largelanguagemodelsllms,
      title={{Large Language Models (LLMs) for Requirements Engineering (RE): A Systematic Literature Review}}, 
      author={Mohammad Amin Zadenoori and Jacek Dąbrowski and Waad Alhoshan and Liping Zhao and Alessio Ferrari},
      year={2025},
      eprint={2509.11446},
      archivePrefix={arXiv},
      primaryClass={cs.SE}
}

@misc{leidinger2023languagepromptinglinguisticproperties,
      title={The language of prompting: What linguistic properties make a prompt successful?}, 
      author={Alina Leidinger and Robert van Rooij and Ekaterina Shutova},
      year={2023},
      eprint={2311.01967},
      archivePrefix={arXiv},
      primaryClass={cs.CL}
}

@inproceedings{wahle-etal-2024-paraphrase,
    title = "Paraphrase Types Elicit Prompt Engineering Capabilities",
    author = "Wahle, Jan Philip  and
      Ruas, Terry  and
      Xu, Yang  and
      Gipp, Bela",
    booktitle = "Empirical Methods in Natural Language Processing",
    year = "2024"
}

@inproceedings{long-etal-2025-makes,
    title = "What Makes a Good Natural Language Prompt?",
    author = "Long, Do Xuan  and
      Dinh, Duy  and
      Nguyen, Ngoc-Hai  and
      Kawaguchi, Kenji  and
      Chen, Nancy F.  and
      Joty, Shafiq  and
      Kan, Min-Yen",
    booktitle = "63rd Annual Meeting of the Association for Computational Linguistics",
    year = "2025",
    ISBN = "979-8-89176-251-0"
}

@inproceedings{cao-etal-2026-style,
    title = "Out of Style: {RAG}{'}s Fragility to Linguistic Variation",
    author = "Cao, Tianyu  and
      Bhandari, Neel  and
      Yerukola, Akhila  and
      Asai, Akari  and
      Sap, Maarten",
    booktitle = "19th Conference of the {E}uropean Chapter of the {A}ssociation for {C}omputational {L}inguistics (Volume 1: Long Papers)",
    year = "2026",
    ISBN = "979-8-89176-380-7"
}

@inproceedings{van2004linguistic,
  title={Linguistic profiling for authorship recognition and verification},
  author={Van Halteren, Hans},
  booktitle={42nd Annual Meeting of the Association for Computational Linguistics},
  pages={199--206},
  year={2004}
}

@inproceedings{dell2011read,
  title={READ--IT: Assessing readability of Italian texts with a view to text simplification},
  author={Dell’Orletta, Felice and Montemagni, Simonetta and Venturi, Giulia},
  booktitle={Proceedings of the second workshop on speech and language processing for assistive technologies},
  pages={73--83},
  year={2011}
}

@article{malmasi2018native,
  title={Native language identification with classifier stacking and ensembles},
  author={Malmasi, Shervin and Dras, Mark},
  journal={Computational Linguistics},
  volume={44},
  number={3},
  pages={403--446},
  year={2018},
  publisher={MIT Press One Rogers Street, Cambridge, MA 02142-1209, USA journals-info~…}
}

@article{futrell2015large,
  title={Large-scale evidence of dependency length minimization in 37 languages},
  author={Futrell, Richard and Mahowald, Kyle and Gibson, Ed.},
  journal={National Academy of Sciences},
  volume={112},
  number={33},
  pages={10336--10341},
  year={2015}
}

@book{halliday1994writing,
  title={Writing Science: Literacy And Discursive Power},
  author={Halliday, Michael Alexander Kirkwood and Martin, JR},
  year={1994},
  publisher={CRC Press}
}

@inproceedings{zhuo-etal-2024-prosa,
    title = "{P}ro{SA}: Assessing and Understanding the Prompt Sensitivity of {LLM}s",
    author = "Zhuo, Jingming  and
      Zhang, Songyang  and
      Fang, Xinyu  and
      Duan, Haodong  and
      Lin, Dahua  and
      Chen, Kai",
    booktitle = "Findings of the Association for Computational Linguistics: EMNLP 2024",
    year = "2024",
    pages = "1950--1976"
}

@InProceedings{Han2026,
author="Han, Shuo
and Tan, Tao
and Miao, Yuantian
and Chen, Xiao
and Sun, Nan",
title="Prompting Instability: An Empirical Study of LLM Robustness in Code Vulnerability Detection",
booktitle="AI 2025: Advances in Artificial Intelligence",
year="2026",
pages="233--245",
}

@article{ClelandHuang2007,
  author  = {Cleland-Huang, Jane and Settimi, Raffaella and Zou, Xuchang and Solc, Peter},
  title   = {Automated Classification of Non-Functional Requirements},
  journal = {Requirements Engineering},
  year    = {2007},
  volume  = {12},
  number  = {2},
  pages   = {103--120},
  issn    = {1432-010X}
}

@dataset{PROMISE,
  author    = {Cleland-Huang, Jane and Mazrouee, Sepideh and Liguo, Huang and Port, Dan},
  title     = {nfr},
  year      = {2007},
  publisher = {Zenodo},
  doi       = {10.5281/zenodo.268542},
  url       = {https://doi.org/10.5281/zenodo.268542}
}

\appendix

\end{document}